\documentclass[aps,prb,twocolumn,showpacs,floatfix]{revtex4}
\usepackage{amsmath}
\usepackage{amssymb}
\usepackage{bm}
\usepackage{graphicx}

\begin{document}


\title
{Decoherence of Excitons in Multichromophore Systems: \\Thermal
Line Broadening and Destruction of Superradiant Emission}

\author{D.\ J.\ Heijs}

\author{V.\ A.\ Malyshev}
\thanks{On leave from S.I. Vavilov State Optical Institute,
Birzhevaya Linia 12, 199034 Saint-Petersburg, Russia.}

\author{J.\ Knoester}
\affiliation{Institute for Theoretical Physics and  Materials
Science Centre, University of Groningen, Nijenborgh 4, 9747 AG
Groningen, The Netherlands}

\date{\today}

\begin{abstract}
We study the temperature-dependent dephasing rate of excitons in
chains of chromophores, accounting for scattering on static
disorder as well as acoustic phonons in the host matrix. From this
we find a powerlaw temperature dependence of the absorption line
width, in excellent quantitative agreement with experiments on dye
aggregates. We also propose a relation between the line width and
the exciton coherence length imposed by the phonons. The results
indicate that the much debated steep rise of the fluorescence
lifetime of pseudo-isocyanine aggregates above 40 K results from
the fact that this coherence length drops below the localization
length imposed by static disorder.
\end{abstract}


\maketitle

A wide variety exists of materials consisting of interacting
chromophores; examples are molecular aggregates, conjugated
polymers, natural light-harvesting systems, and arrays of quantum
dots. Lately, such multichromophore systems have received much
attention for their collective optical
properties,\cite{Meinardi03,Lim04,Hoffman03,Bednarz03} efficient
and controlled energy transport
processes,\cite{Schindler04_Crooker02_Jang04_Fleming04} the
occurrence of strong quantum entanglement of collective
chromophore states\cite{Reina00,Hettich02} and their possible
application in schemes for quantum
computation.\cite{Reina00,Lovett03} All these properties
sensitively depend on the temporal and spatial energy and phase
relaxation of the chromophores' collective excitations, which make
such relaxation processes a topic of broad interest and
importance. The nature of the collective excitations and their
dynamics are governed by the complicated interplay of their
scattering on static disorder (e.g., inhomogeneity in the
excitation energy of the individual chromophores) and on dynamic
degrees of freedom (vibrations). A variety of experimental
techniques are used to investigate the resulting intricate
dynamics. The theoretical study of the combined effects of static
disorder and a heat bath on exciton decoherence and optical
response remains relatively unexplored.\cite{Mukamel04}

Recently, a model of weakly-localized Frenkel excitons coupled to
acoustic phonons in the host medium proved to yield an excellent
description of the nonmonotonous temperature dependence of the
fluorescence Stokes shift as well as the linear temperature scaling
of the radiative lifetime of aggregates of the dye
3,3'-bis(sulfopropyl)-5,5'-dichloro-9-ethylthiacarbo-cyanine
(THIATS) measured between 0 and 100 K.\cite{Bednarz03} More direct
insight into the interplay of static disorder and homogeneous
dephasing, and thus a more stringent test for the model, is
provided by temperature dependent absorption, photon echo, and
hole burning measurements. Such measurements, over the interval
0-300 K, have been performed for the prototypical J-aggregates of
the dye pseudo-isocyanine
(PIC).\cite{deBoer87,Hirschmann89,Fidder90,Fidder95,Renge97} No
unique interpretation of the thermal dephasing and line broadening
in multichromophore systems has emerged from these studies:
activated processes involving several optical phonons of the
aggregate have been suggested as
mechanism,\cite{Hirschmann89,Fidder90,Fidder95} but also power-law
broadening due to host vibrations.\cite{Renge97} Another unsolved
issue is the temperature dependence of the fluorescence lifetime
at elevated temperatures. For aggregates of PIC-Br, its steep rise
above 40 K has been only explained using a two-dimensional
aggregate model,\cite{Potma98} which contradicts the generally
accepted chain-like geometry of these systems in
solution.\cite{Berlepsch00}

In this Letter, we revisit the model of excitons in a linear chain
with static energy disorder, coupled to acoustic phonons in the
host. We show that the dephasing rates calculated within this
model give an excellent explanation of the temperature dependent
absorption measurements for PIC aggregates. Furthermore, for the
strong scattering regime we introduce a relation between the
exciton coherence length and the dephasing rates and demonstrate
that this explains the steep rise of the fluorescence lifetime.

Our model consists of a linear Frenkel chain of $N$ two-level
chromophores ($n=1,\ldots,N$), with uncorrelated Gaussian disorder
of standard deviation $\sigma$ in the chromophore energies
$\varepsilon_n$ and resonant dipole-dipole interactions between
chromophores $n$ and $m$ given by $J_{nm}=-J/|n-m|^3$, $J > 0$
being the nearest-neighbor interaction. All transition dipoles are
parallel. The exciton eigenstates (labeled $\nu$) follow from
diagonalizing the $N \times N$ matrix with the $\varepsilon_n$ as
diagonal elements and the $J_{nm}$ as off-diagonal ones. They have
energy $E_{\nu}$, and site-amplitudes $\varphi_{\nu n}$; the
disorder leads to their localization on segments of the chain. The
model includes on-site scattering of excitons on a phonon bath. If
the exciton-phonon interaction is not too strong, it induces
scattering from exciton state $\nu$ to state $\mu$ with a rate
that may be obtained from Fermi's Golden rule (see, e.g.
Ref.~\onlinecite{Bednarz03}):
\begin{equation}
\label{W}
    W_{\mu\nu} =   {\cal F}(|\omega_{\mu\nu}|)G(\omega_{\mu\nu})
    \sum_{n=1}^N \varphi_{\mu n }^2 \varphi_{\nu n}^2 \ .
\end{equation}
Here, $\omega_{\mu\nu} = E_\mu-E_\nu$ and ${\cal F}(\omega)=2 \pi
\sum_q |V_q|^2 \delta(\omega-\omega_q)$, the one-phonon spectral
density ($\omega_q$ is the energy of a phonon in mode $q$ and
$V_q$ characterizes the coupling of this mode to the excitons; we
set $\hbar=1$). Furthermore, $G(\omega)=n(\omega)$ if $\omega>0$
and $G(\omega)=1+n(-\omega)$ if $\omega<0$, with
$n(\omega)=[\exp(\omega/k_B T)-1]^{-1}$, the mean thermal
occupation number of a phonon mode of energy $\omega$.

While in Ref.~\onlinecite{Bednarz03} the $W_{\mu \nu}$ were only
used to study the intraband exciton relaxation and the resulting
fluorescence kinetics, they also govern the temperature dependent
dephasing of the excitons, and hence their homogeneous line
widths. Specifically, the thermal dephasing rate of the $\nu$th
exciton state is given by $\Gamma_{\nu} \equiv \frac{1}{2}
\sum_\mu W_{\mu \nu}$ and depends on temperature through the
$n(\omega)$. In contrast to the fluorescence kinetics, the
temperature dependence of the dephasing rates and absorption line
width sensitively depends on the form assumed for the spectral
density. Here, we will assume that ${\cal F}(\omega) = W_0
(\omega/J)^3$. This form is appropriate for acoustic phonons, for
which the density of states (DOS) scales as $\omega^2$ (Debye
behavior) and $|V_q|^2$ scales like $\omega$; the remaining factor
$W_0$ is a free parameter in the model. As we will see, the
$\omega^3$ scaling gives a natural explanation for the line width
measurements on linear dye aggregates. Similar good agreement
between theory and experiment is found if we allow for
fluctuations of ${\cal F}(\omega)$ around an average $\omega^3$
scaling, but the agreement is lost by changing to a power
different from 3.\cite{Heijs05}

Using the above model, we simulated the absorption spectrum for an
ensemble of chains. The expression for this spectrum reads:
\begin{equation}
\label{A}
    A(E) = \frac{1}{N} \left\langle \sum_\nu  \frac{F_\nu}{\pi} \,\,
    \frac{\Gamma_\nu + \gamma_\nu/2}{(E - E_\nu)^2 +
    (\Gamma_\nu + \gamma_\nu/2)^2}\right\rangle \ .
\end{equation}
where $F_\nu = (\sum_n \varphi_{\nu n})^2$ is the dimensionless
oscillator strength of the $\nu$th exciton state, $\gamma_\nu =
\gamma_0 F_\nu$ is its radiative rate ($\gamma_0$ is the emission
rate of a single chromophore) and the angular brackets denote
averaging over the static disorder in the $\varepsilon_n$. The
resulting line width $\Delta(T)$ is determined as the full width
at half maximum (FWHM) of the simulated spectrum. We used chain
lengths of 250-1000 chromophores and averaged over 1000-400,000
disorder realizations, depending on parameter choices.

In the absence of disorder and for long chains with
nearest-neighbor interactions only, we found that $\Delta(T)
\propto T^{3.5}$ (neglecting the radiative contribution, which is
allowed unless $T$ is very low). The explanation is simple: for
$\sigma=0$, the lowest ($\nu = 1$) exciton state dominates the
spectrum. The above scaling then simply results from summing the
$W_{\mu,\nu=1}$ over all higher lying states $\mu$. The $\omega^3$
dependence of ${\cal F}(\omega)$, combined with the $\omega^{-1/2}$
dependence of the exciton DOS near the bottom of the
one-dimensional band, upon integration yields the $T^{3.5}$
behavior. It turns out that the long-range dipolar interactions
slightly change the power from 3.5 to 3.85.~\cite{Heijs05}

\begin{figure}[ht]
\includegraphics[width= \columnwidth,clip]{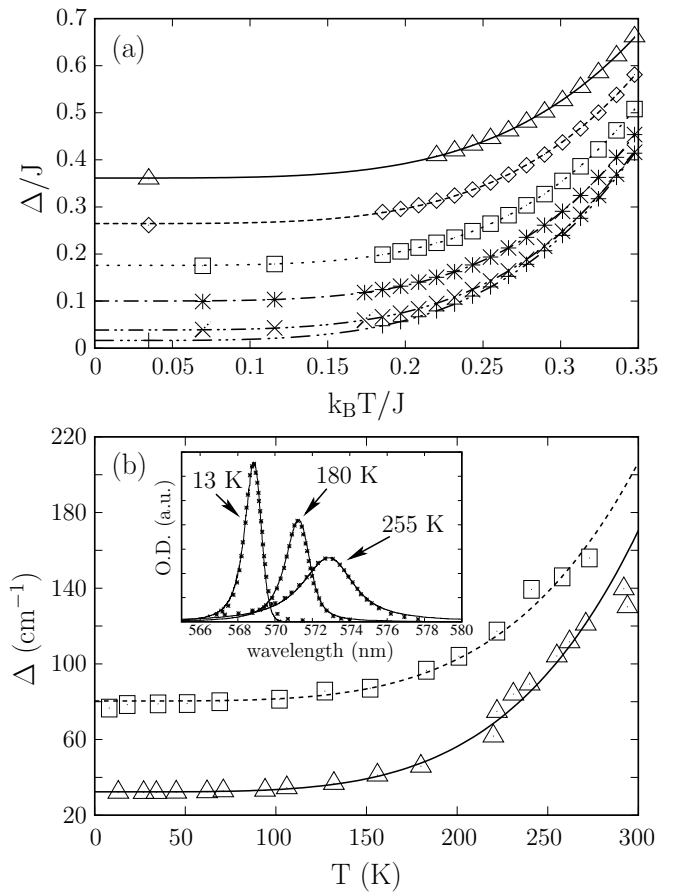}
\caption{(a) Calculated temperature-dependent width of the
absorption spectrum (symbols) and corresponding fits to
Eq.~(\ref{scaling}) (lines) for six values of the degree of
disorder $\sigma/J = 0.05$, 0.1, 0.2, 0.3, 0.4, 0.5 (bottom to
top). The fit parameters $(a,p)$ take the values (1.1, 4.0), (1.2,
4.2), (1.3, 4.3), (1.2, 4.3), (1.2, 4.3), and (1.0, 4.1),
respectively. $W_0/J=25$ and $\gamma_0/J=1.5 \times 10^{-5}$. (b)
Comparison of theory (curves) and experiment (symbols) for the
absorption line width as a function of temperature for aggregates
of PIC-Cl (triangles) and PIC-F (squares). The model parameters
are given in the text. The inset compares the line shape
calculated for PIC-Cl aggregates (curves) to experiment (symbols)
for three temperatures. We note that the thermal shift of the
absorption band is not accounted for in our theory; the calculated
spectra were shifted by hand to facilitate comparison of the line
shape. Experimental data were taken from
Ref.~\onlinecite{Renge97}.} \label{renge_abs}
\end{figure}

If we include disorder, the spectral width at zero temperature,
$\Delta(0)$, is dominated by the inhomogeneous width, which for
linear aggregates scales like $\sigma^{4/3}$.\cite{Schreiber82}
From extensive simulations we have found that in the presence of
disorder the thermal broadening follows a power-law:
\begin{equation}
\label{scaling}
    \Delta(T) = \Delta(0) + aW_0(k_B T/J)^{p} \ ,
\end{equation}
where $a$ and $p$ depend only weakly on the parameters $\sigma$
and $W_0$. The scaling relation Eq.~(\ref{scaling}) implies that,
although the spectrum results from a distribution of exciton
states with different dephasing rates, the total width may be
interpreted as the sum of an inhomogeneous width, $\Delta(0)$, and
a dynamic contribution. For six $\sigma/J$ values in the interval
[0.05, 0.5] (taking $W_0/J=25$), our simulated data are presented
as symbols in Fig.~\ref{renge_abs}(a), together with fits to
Eq.~(\ref{scaling}). The fit parameters are given in the caption.
The power $p$ is seen to lie in a narrow range, $p=4.2 \pm 0.2$,
which is close to the disorder-free value 3.85. The reason for
this small effect of disorder is that the dynamic width is
governed by scattering from the optically dominant DOS tail states
to high-lying intra-band states, whose DOS is modified only
slightly by disorder. The small increase of $p$ primarily results
from the possibility to scatter further downward in the tail of
the disordered exciton band.

A scaling relation like Eq.~(\ref{scaling}) was also found
experimentally by Renge and Wild\cite{Renge97} for aggregates of
PIC with counter ions Cl$^-$ and F$^-$. Fits of their data,
presented in Fig.~\ref{renge_abs}(b), demonstrate good agreement
between theory and experiment. These fits were obtained using the
accepted values of $J = 600$ cm$^{-1}$ and $\gamma_0 = 2.7\times
10^{8}$ s$^{-1}$ ($=1.5\times 10^{-5} J$) for PIC and taking the
fit parameters $\sigma/J=0.128$ and $W_0/J=16.0$ for PIC-Cl and
$\sigma/J=0.249$ and $W_0/J=15.9$ for PIC-F. In the inset we show
(for PIC-Cl) that not only the width, but also the experimental
line shape as a function of temperature is recovered very well by
our model.

\begin{figure}[ht]
\includegraphics[width= \columnwidth,clip]{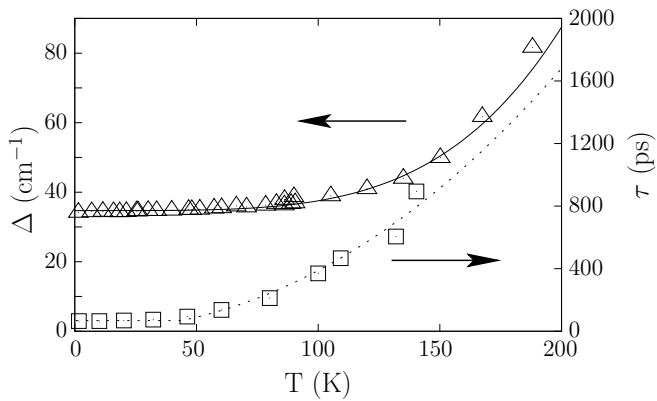}
\caption{Calculated and measured line width and fluorescence
lifetime for aggregates of PIC-Br as a function of temperature.
Line width: triangles (measured\cite{Fidder90}) and solid
(calculated); lifetime: squares (measured\cite{Fidder95}) and
dashed (calculated). } \label{bromide}
\end{figure}

We next address PIC-Br aggregates, for which the absorption line
width as well as the fluorescence lifetime were measured up to 190
K (triangles and squares, respectively, in Fig.~\ref{bromide}). We
first discuss the width, for which it turned out that our model
also for this counter-ion gives a very good fit to experiment
(solid line in Fig.~\ref{bromide}, fit parameters
$\sigma/J=0.135$, $W_0/J=32.5$). It should be noted that the same
data have previously been fitted assuming activated behavior,
supposedly resulting from scattering on optical phonons of the
aggregate.\cite{Fidder95} In order to describe the strong increase
of the line width for large temperatures, such a fit requires both
a high activation energy (several hundreds of cm$^{-1}$) and a
large pre-exponential factor. The pre-factor is directly related
to the exciton-vibration coupling constant. A simple perturbative
treatment shows that the thus estimated coupling constant should
be so large (few thousand cm$^{-1}$'s) that a strong polaron effect
is to be expected. This is inconsistent with the generally
accepted excitonic nature of the optical excitations in cyanine
aggregates.

At low temperature, the fluorescence lifetime is determined by a
competition between spontaneous emission and intraband relaxation
of the weakly localized excitons, and may be analyzed using a
master equation for these populations.\cite{Bednarz03} While
successful for THIATS aggregates up to 100 K, this approach turns
out not to explain the steep rise of the measured lifetime for
PIC-Br above about 40 K (cf.~Fig.~\ref{bromide}). In fact, using
the above fit parameters, the model only yields a doubling of the
lifetime from 0 to 150 K, which is an order of magnitude too
small. Other approaches have seen similar problems in explaining
the measurements;\cite{Fidder90,Spanonote} the closest (far from
perfect) agreement has been found using a two-dimensional
aggregate model.\cite{Potma98} Here we argue that the problem
arises from phonon-induced decoherence of the excitons inside
their localization segments.

If we consider weak exciton-phonon scattering in homogeneous chains
($\sigma=0$), the exciton coherence length (or mean-free path),
$N_{\mu}^\mathrm{coh}$, is much larger than its wavelength. In
other words, the uncertainty in the exciton's quasi-momentum,
$\delta K \sim 1/N_{\mu}^\mathrm{coh}$, is much smaller than the
quasi-momentum itself. Due to the quadratic dispersion relation
near the band bottom, $E_\mu = 2J - JK^2$, $\delta K$ results in
an energy uncertainty $\delta E_\mu = 2JK \delta K \sim
JK/N_{\mu}^\mathrm{coh}$. Identifying $\delta E_\mu$ with the
exciton's homogeneous width $\Gamma_\mu$, we thus obtain an
estimate for the coherence length: $N_{\mu}^\mathrm{coh} \sim
JK/\Gamma_\mu$. In the strong scattering regime,
$N_{\mu}^\mathrm{coh}$ becomes smaller than the exciton wavelength
and the quasi-momentum is not a good quantum number anymore:
$\delta K \sim K$. Hence $\delta E_\mu \sim J (\delta K)^2 \sim
J/(N_{\mu}^\mathrm{coh})^2$, leading to the estimate
$N_{\mu}^\mathrm{coh} \sim (J/\Gamma_\mu)^{1/2}$.

We now address this issue for the localized band edge states in
the presence of disorder. At low temperature [$k_BT \ll
\Delta(0)$], the coherence length $N_{\mu}^\mathrm{coh}$
associated with the exciton-phonon scattering is much larger than
the localization length $N_\mu^\mathrm{loc}$; hence the latter
plays the role of the coherence length and determines the optical
dynamics. The main consequence of the exciton-phonon scattering in
this regime is the intraband relaxation (equilibration) described
above already.~\cite{Bednarz03} Upon increasing $T$,
$N_{\mu}^\mathrm{coh}$ decreases and at some temperature $T_0$
becomes comparable with $N_\mu^\mathrm{loc}$. In other words,
excitons start to scatter {\it within} localization segments. For
$T > T_0$, the coherence length $N_{\mu}^\mathrm{coh}$, rather than
the localization length, governs the optical dynamics. It is to be
stressed that when $T$ surpasses $T_0$, the optically dominant
states immediately fall in the strong scattering regime, because
they resemble standing waves without nodes with a typical
"wavelength" on the order of
$N_\mu^\mathrm{loc}$.~\cite{Malyshev95}

The cross-over temperature $T_0$ separates two different regimes
for the fluorescence lifetime as a function of temperature. This
lifetime is inversely proportional to the number of coherently
emitting chromophores. At low temperature, this number should be
identified with the typical localization length of the optically
dominant exciton states, $N^*$; at high temperatures, it is to be
identified with the $T$-dependent coherence length,
$N^\mathrm{coh}(T)$. Denoting $\tau_0=1/\gamma_0$ as the
single-chromophore lifetime, the essential behavior of the
lifetime is thus given by:
\begin{eqnarray}
\label{tau}
    \tau(T) = \left\{
\begin{array}{lr}
    \tau_0/N^* \ , &\quad T <  T_0 \ ,
\\
\\
    \tau_0/N^\mathrm{coh}(T) = \tau_0 b\sqrt{\Gamma(T)/J} \ ,
    &\quad T > T_0 \ ,
\end{array}
\right.
\end{eqnarray}
where in the estimate for the coherence length we used for the
typical dephasing rate the effective homogeneous width
$2\Gamma(T)\equiv \Delta(T)-\Delta(0)$ and $b$ is a parameter of
order unity. Here, the value of $T_0$ is determined from the
condition $b\sqrt{\Gamma(T_0)/J}=1/N^*$.

We used this approach to analyze the fluorescence data for PIC-Br.
We took $N^*=54$, as follows from the low-temperature lifetime
$\tau(0) = 68$ ps in combination with $\tau_0 = 3.7$
ns,~\cite{Fidder90,Fidder95}, and which also is consistent with
the participation ratio\cite{Schreiber82} of 58 calculated for
$\sigma/J=0.135$. The resulting very good fit, using the
$\Delta(T)$ calculated above, is seen as dashed line in
Fig.~\ref{bromide}. Here we used $b=2.2$ as only fit parameter,
which gave $T_0=44$ K.

Two comments are in place. First, in the low-temperature regime
our fit does not account for population relaxation, which for
growing $T$ increases the lifetime from the value $\tau_0/N^*$ due
to thermal population of dark states. Using the master-equation
approach of Ref.~\onlinecite{Bednarz03}, we found that this effect
for PIC-Br leads to a small ($\sim$ 20 ps) linear growth of the
lifetime between $T=0$ and $T=T_0$. Second, a scaling of the
lifetime with the line width was also suggested by Feldmann {\em et
al.}\cite{Feldmann87} for quantum wells. Translating their
arguments to a one-dimensional system, one obtains $\tau(T)=
(\tau_0/\pi)\sqrt{\Delta(T)/J}$, which in the high-temperature limit
agrees with our result, except that the pre-factor $b$ is reduced
considerably. The reason for this difference is that the argument
of Ref.~\onlinecite{Feldmann87}, based on counting the number of
states within the absorption width, does not relate to the
coherence length, but rather to population redistribution.

In conclusion, a model that considers weakly-localized excitons
which scatter on acoustic host phonons, yields power-law thermal
line broadening that is in excellent agreement with experiments on
dye aggregates. In addition, a new method to relate the exciton
coherence length within the strong-scattering regime to the
homogeneous line width, gives good agreement with temperature
dependent fluorescence lifetime data. Our work shows that, in
contrast to earlier claims, a one-dimensional model for PIC
aggregates suffices to explain all experimental data. We believe
that both the on-site (diagonal) disorder and the one-phonon
scattering on host vibrations with a Debye-like spectral density
($\propto \omega^3$) are the essential and physically sound
ingredients to reach this conclusion. It is known that
off-diagonal disorder yields a symmetric low-temperature
absorption profile,~\cite{Fidder91} while two-phonon scattering
(including pure dephasing) results in power-law broadening with a
power that is about twice as large as the values of $p$ obtained
for one-phonon scattering.~\cite{Heijs05} Both facts are
inconsistent with experimental data. Finally, off-site scattering
on acoustic phonons is also characterized by a spectral density
proportional to $\omega^3$ (see, e.g. the work cited in
Ref.~\onlinecite{Spanonote}); therefore its inclusion would not
change the power $p=4.2$, but only alter the microscopic meaning
of the phenomenological scattering parameter $W_0$ to a sum of two
(on-site and off-site) scattering parameters.

\end{document}